\documentclass[a4paper,fleqn]{cas-sc}

\usepackage{amsmath, amssymb, amsthm}
\usepackage{mathtools}

\usepackage{enumitem}

\usepackage{graphicx}

\usepackage{xcolor}

\usepackage{glossaries}

\usepackage{makecell}

\usepackage{tikz}

\usepackage{blindtext}

\usepackage{bm}
\usepackage{algorithmic}
\usepackage{algorithm}
\usepackage[]{subfig}
\usepackage{booktabs}
\usepackage{siunitx}

\usepackage[numbers]{natbib}

\usepackage{xr}

\usepackage{hyperref}
\usepackage[capitalise]{cleveref}

\usepackage[pagewise]{lineno}

\graphicspath{{./img/}{./plots/}}

\newlength{\imgWidth}
\definecolor{my_red}{rgb}{0.5 0 0}
\definecolor{my_blue}{rgb}{0 0.3529 0.6627}
\definecolor{my_orange}{rgb}{0.9608, 0.5020, 0.1451}
\definecolor{my_green}{rgb}{0, 0.38, 0}

\definecolor{my_red}{rgb}{0.9 0 0.1} \definecolor{my_blue}{rgb}{0 0.3529 0.6627} \definecolor{my_orange}{rgb}{0.925, 0.396, 0} \definecolor{my_green}{rgb}{0.6, 0.753, 0} 

 \DeclareMathOperator{\E}{\mathsf{E}}
\DeclareMathOperator{\Var}{\mathsf{Var}}

\DeclareMathOperator*{\argmin}{\mathrm{arg\,min}}
\DeclareMathOperator{\Tr}{\mathrm{Tr}}

\newcommand{\given}{\,|\,}
\newcommand{\bgiven}{\,\big|\,}

\newcommand{\Nmax}{N}

\newcommand{\Hyp}{\mathrm{H}}
\newcommand{\param}{\boldsymbol\theta}
\newcommand{\paramRV}{\boldsymbol\Theta}

\newcommand{\stat}[1][n]{\mathbf{t}_{#1}}

\newcommand{\obsFSS}[1][N]{\mathbf{x}_{1:#1}}
\newcommand{\RVfss}[1][N]{\mathbf{X}_{1:#1}}

\newcommand{\obsSingle}{\mathbf{x}}
\newcommand{\RVsingle}{\mathbf{X}}
\newcommand{\obsSeq}[1][n]{\mathbf{x}_{1:#1}}

\newcommand{\policy}{\pi}

\newcommand{\decR}[1][n]{\delta_{#1}}
\newcommand{\decROpt}[1][n]{\delta_{#1}^\star}
\newcommand{\stopR}[1][n]{\Psi_{#1}}
\newcommand{\stopAt}[1][n]{\Phi_{#1}}
\newcommand{\rl}{\tau}
\newcommand{\est}[2][n]{\hat{\boldsymbol\theta}_{#2,#1}}

\newcommand{\estOpt}[2][n]{\hat{\boldsymbol\theta}^\star_{#2,#1}}

\newcommand{\detErr}[1][i]{\alpha^{#1}}
\newcommand{\estErr}[1][i]{\beta^{#1}}
\newcommand{\detConstr}[1][i]{\bar{\alpha}^{#1}}
\newcommand{\estConstr}[1][i]{\bar{\beta}^{#1}}
\newcommand{\detCost}[1][i]{\lambda_{#1}}
\newcommand{\estCost}[1][i]{\mu_{#1}}

\newcommand{\dInt}{\mathrm{d}}

\newcommand{\contCost}[1][n]{d_{#1}}

\newcommand{\stopRopt}[1][n]{\Psi_{#1}^\star}
\newcommand{\stopRao}[1][n]{\Psi_{#1}^\circ}

\newcommand{\indd}[1]{\mathbf{1}_{#1}} \newcommand{\ind}[1]{\indd{\{#1\}} }

\newcommand{\lr}{\eta}
\newcommand{\lrGLRT}{\eta^\text{GLR}}
\newcommand{\lrBayes}{\eta^\text{B}}

\newcommand{\thNP}{\lambda^\text{NP}}

\newlength{\imgWidthSingle}
\newlength{\imgWidthDouble}
\DeclareRobustCommand\dashed{\tikz[baseline=-0.6ex]\draw[thick,dashed] (0,0)--(0.54,0);}

\newacronym{apo}{APO}{asymptotically pointwise optimal}
\newacronym{ao}{AO}{asymptotically optimal}
\newacronym{awgn}{AWGN}{additive white Gaussian noise}
\newacronym{iid}{iid}{independent and identically distributed}
\newacronym{kl}{KL divergence}{Kullback-Leibler divergence}
\newacronym{MAE}{MAE}{mean absolute error}
\newacronym{mmse}{MMSE}{minimum mean-squared error}
\newacronym{MSE}{MSE}{mean-squared error}
\newacronym{ML}{ML}{maximum likelihood}
\newacronym{msprt}{MSPRT}{Matrix Sequential Probability Ratio Test}
\newacronym{pdf}{pdf}{probability density function}
\newacronym{sprt}{SPRT}{sequential probability ratio test}
\newacronym{gsprt}{GSPRT}{generalized sequential probability ratio test}
\newacronym{qam}{QAM}{quadrature amplitude modulation}
\newacronym{bfgs}{BFGS}{Broyden–Fletcher–Goldfarb–Shanno}
\newacronym{fim}{FIM}{Fisher's information matrix}
\newacronym{fss}{FSS}{fixed sample-size}
\newacronym{lr}{LR}{likelihood ratio}
\newacronym{ts}{TS}{two-step}
\newacronym{lp}{LP}{linear program}

\newacronym{FC}{FC}{fusion center}

\newacronym{GLRT}{GLRT}{Generalized Likelihood Ratio Test}
\newacronym{NP}{NP}{Neyman-Pearson}
\newacronym{MAP}{MAP}{maximum a posteriori}

\newacronym{JDE}{JDE}{joint detection and estimation}
\newacronym{SJDE}{SJDE}{sequential joint detection and estimation}

\newacronym{ci}{CI}{Consensus+Innovations}

\newacronym{CRB}{CRB}{Cram\'{e}r--Rao bound}

\graphicspath{{./imgs/}}

\makeatletter
\newlength{\hb@colwidth}
\def\makeheaderbox{
	\noindent\textcolor{black}{\rule{\textwidth}{1pt}}
	\par\vskip2pt\noindent
	\begin{minipage}{\textwidth}
		\centering
		\setlength{\tabcolsep}{0pt}
		\setlength{\extrarowheight}{0pt}
		\setlength{\hb@colwidth}{0.16666667\textwidth}
		\renewcommand{\arraystretch}{2}
		\begin{tabular}{
				p{\hb@colwidth}
				>{\columncolor[gray]{0.95}}p{4\hb@colwidth}
				p{\hb@colwidth}
			}
			\parbox[c][\hb@colwidth][c]{\hb@colwidth}{
				\centering
				\includegraphics[height=0.9\hb@colwidth, keepaspectratio]{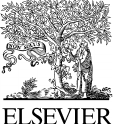}
			} & 
			\parbox[c][\hb@colwidth][c]{4\hb@colwidth}{
				\centering
				{\fontsize{8}{10}\selectfont Contents lists available at \href{https://www.elsevier.com/locate/fi}{ScienceDirect}}\par
				\vspace{15pt}
				{\fontsize{16}{18}\selectfont Journal of the Franklin Institute}\par
				\vspace{12pt}
				{\fontsize{8}{10}\selectfont journal homepage: \href{https://www.elsevier.com/locate/fi}{www.elsevier.com/locate/fi}}
			} & 
			\parbox[c][\hb@colwidth][c]{\hb@colwidth}{
				\centering
				\includegraphics[height=\hb@colwidth, keepaspectratio]{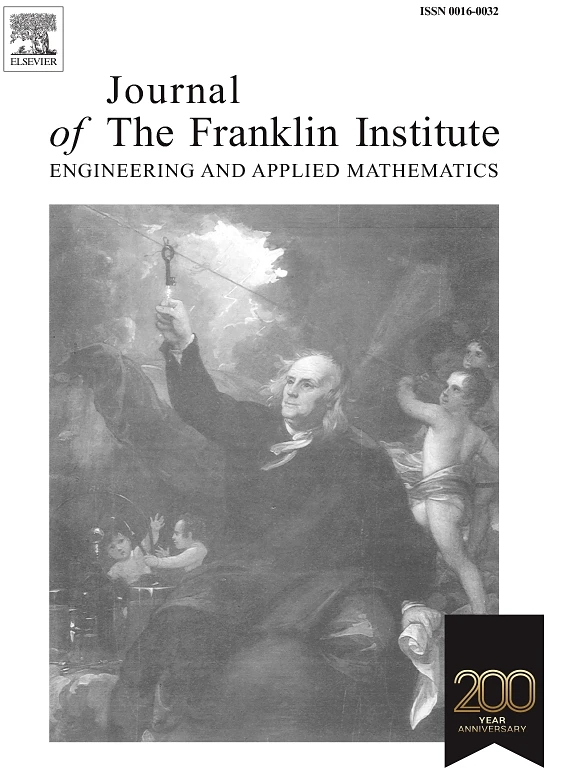}
			} \\
		\end{tabular}
	\end{minipage}
	\par\vskip2pt
	\noindent\textcolor{black}{\rule{\textwidth}{3pt}}
	\par\vskip18pt
}

\let\originalmaketitle\maketitle
\def\maketitle{
	\makeheaderbox
	\vspace{2pt}
	\begin{center}{\Large\bfseries \emph{Bicentennial Special Issue Invited Paper}}
    \end{center}\par
	\vspace{10pt}
	\originalmaketitle
}
\makeatother

\crefname{equation}{}{}

\definecolor{decColorH1}{rgb}{0.0, 0.44, 1.0}
\definecolor{decColorH2}{rgb}{1.0, 0.01, 0.24}
\definecolor{decColorH3}{rgb}{1 0.87 0}
\definecolor{decColorH4}{rgb}{0.0, 0.42, 0.24}

\colorlet{contColor}{gray!40}

\begin{document}

\let\WriteBookmarks\relax
\def\floatpagepagefraction{1}
\def\textpagefraction{.001}
\shorttitle{Journal of the Franklin Institute}
\shortauthors{D. Reinhard et~al.}

\title [mode = title]{(Sequential) Joint Detection and Estimation: Classic Results and New Directions}

\author[1]{Dominik Reinhard}[
                        auid=000,bioid=1,
orcid=0000-0002-6250-6221
                       ]

\ead{reinhard@spg.tu-darmstadt.de}

\affiliation[1]{organization={Signal Processing Group, Technische Universit\"at Darmstadt},
                addressline={Merckstra\ss{}e 25}, 
                city={Darmstadt},
                country={Germany}}

\author[1]{Abdelhak M. {Zoubir}}[type=editor,
                        auid=001,bioid=2,
                        orcid=0000-0002-4409-7743
]
\cormark[1]   
\ead{zoubir@spg.tu-darmstadt.de}

\cortext[cor1]{Corresponding author}

\begin{abstract}
We provide an overview of the problem of jointly testing two hypotheses and estimating a parameter of the selected model.
Such problems arise in a variety of applications.
First, we present a conceptual introduction to suboptimal and optimal procedures for joint detection and estimation.
A numerical example illustrates the advantages of the optimal procedure over suboptimal ones.
Next, we discuss how more advanced problem formulations affect the presented results.
The second part covers joint detection and estimation in a sequential framework.
First, we provide an introduction to sequential analysis through sequential hypothesis testing.
Then, suboptimal and optimal sequential procedures for joint detection and estimation are discussed.
A numerical example shows the advantages of optimal sequential procedures over suboptimal and optimal sequential procedures with a fixed number of samples.
The third part discusses open problems and future research directions in joint detection and estimation.
\end{abstract}


\maketitle

\glsresetall
\section{Introduction}
Hypothesis testing and parameter estimation are two fundamental problems in signal processing and statistics.
Researchers have focused on these problems for decades, deriving optimal solutions under various scenarios.
However, in many applications one is not solely interested in hypothesis testing or parameter estimation, but rather in the two intrinsically coupled problems.
The main objective of a radar system, for example, is not only to detect the presence of a target, but also to provide an estimate of the position and/or the velocity of the target \cite{Tajer2009,Dong2021,Zhang2025,tajer2010optimal,chen2018impact}.
Often, the problems of hypothesis testing and parameter estimation are considered separately, e.g., one first tries to detect a target and parameter estimation is performed once a target is detected.
However, the outcome of the detector and the outcome of the estimator are both of primary interest.
Considering both sub-problems separately does not necessarily result in an overall optimal performance even if an optimal detector and an optimal estimator are used \cite{moustakides2012joint}.
Hence, the problems of detection and estimation have to be solved \emph{jointly}, which is referred to as \gls{JDE}.

Besides the area of radar signal processing \cite{Tajer2009,Dong2021,Zhang2025,tajer2010optimal}, coupled detection and estimation occur in a wide range of applications.
In communications, for example, the problem of channel estimation naturally occurs besides detection tasks \cite{jan2018iterative,Jiang2020,Viakalo2006}.
In cognitive radio, the secondary user aims to detect the primary user and simultaneously estimate possible interference \cite{Yilmaz2014Sequential}.  
Moreover, the problems of detection and estimation occur in a coupled way in speech processing \cite{momeni2015joint}, biomedical engineering \cite{chaari2012fast,makni2008fully}, changepoint detection \cite{boutoille2010hybrid}, visual  inference \cite{vo2010joint}, industrial plant monitoring \cite{Gianluca2024} or power state monitoring \cite{sihag2018power}.

The problem of \gls{JDE} goes back to the 1960s when it was first investigated for the binary hypothesis case \cite{middleton1968simultaneous}.
Later, it was extended to the problem of joint $M$-ary hypothesis testing and parameter estimation \cite{fredriksen1972simultaneous}.
In the new millennium, the problem of \gls{JDE} regained attention.
Besides fundamental research \cite{Yilmaz2014Sequential,reinhard2018,reinhard2022,yilmaz2015sequential,Yilmaz2016Sequential,Lyu2024,reinhard2026minimax,reinhard2024}, also applications, such as, radar \cite{Tajer2009, Dong2021,Zhang2025, tajer2010optimal,moustakides2012joint}, communications \cite{jan2018iterative,Viakalo2006}, speech processing \cite{momeni2015joint}, (medical) imaging \cite{vo2010joint,makni2008fully,chaari2012fast}, changepoint detection \cite{boutoille2010hybrid} or power engineering \cite{sihag2018power} have been investigated.

\begin{figure}

    \subfloat[Radar beam pattern with strong, but wide peak.\label{fig:sjde_radar_det}]{
    {\setlength{\fboxsep}{0pt}
    \framebox{
    \includegraphics[width=0.27\textwidth,trim={3.5cm 3cm 5cm 2cm},clip]{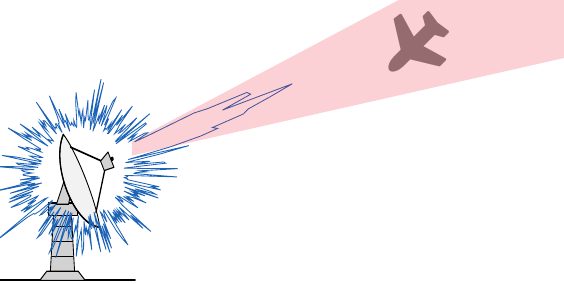}}}}
    \hfill
    \subfloat[Radar beam pattern with narrow, but weak peak.\label{fig:sjde_radar_est}]{
    {\setlength{\fboxsep}{0pt}
    \framebox{
    \includegraphics[width=0.27\textwidth,trim={3.5cm 3cm 0cm 2cm},clip]{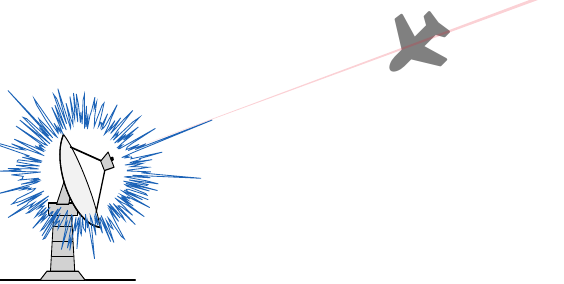}}}}
    \hfill
    \subfloat[Radar beam pattern with strong and narrow peak.\label{fig:sjde_radar_joint}]{
        {\setlength{\fboxsep}{0pt}
    \framebox{
            \includegraphics[width=0.27\textwidth,trim={3.5cm 3cm 0cm 1.85cm},clip]{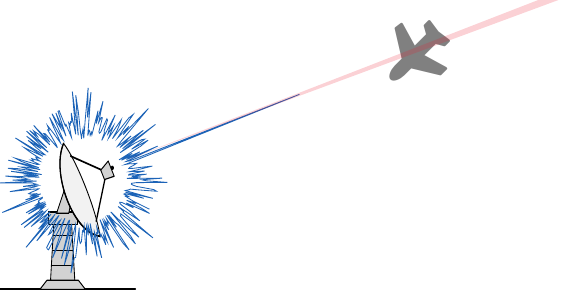}}}}
    
 \caption{Radar scenario in which one wishes to detect a target and estimate its position. The uncertainty about the target position is indicated by the red shaded area.}
 \label{fig:radar_jde}
\end{figure}

Traditionally, inference is based on a fixed number of samples, which is also known as \gls{fss} scenario.
Alternatively, the field of sequential analysis arose in the late 1940s \cite{wald1945sequential,wald1948}.
In sequential inference, the data arrives as streaming data and one stops sampling as soon as one is confident about the phenomenon of interest.
In his pioneering work, Wald first applied this to statistical hypothesis testing, resulting in one of the most widely used sequential tests, the \gls{sprt} \cite{wald1945sequential,wald1948}.
Sequential hypothesis tests can save on average up to $50\,\%$ of the samples compared to a \gls{fss} test with similar performance.
Besides the fact that sequential methods use on average much fewer samples than \gls{fss} methods, the variable sample number in sequential analysis provides an additional important degree of freedom. 
In sequential \gls{JDE}, this allows an individual control of the error probabilities and the estimation error levels and, hence, provides an important and powerful framework.
In a sequential framework, this is of particular importance as detection and estimation sometimes have contradicting requirements on the data.
For illustration, consider again the problem of target detection and localization in radar, for which three different realizations are sketched in \cref{fig:radar_jde}.
In \cref{fig:sjde_radar_det}, the radar beam pattern shows a prominent, but wide peak in the direction of the target.
Based on this data set, a reliable detection is possible, but the wide peak results in a high uncertainty about the position of the target.
The radar pattern in \cref{fig:sjde_radar_est} shows a narrow peak in the direction of the target that is only slightly above the noise floor.
Hence, this results in a precise target localization, but a reliable detection is not possible.
Finally, the radar pattern in \cref{fig:sjde_radar_joint} shows a prominent and narrow peak in the direction of the target, which allows a reliable detection and localization of the target.
Therefore, in a sequential setup, one must take additional samples until the confidence about the underlying hypothesis and the unknown parameter is high enough.

After the first steps towards \gls{SJDE} \cite{buzzi2006,grossi2008}, the problem has been investigated by Y{\i}lmaz \emph{et al.} \cite{yilmaz2015sequential,Yilmaz2014Sequential,Yilmaz2016Sequential}, who aimed to minimize the number of samples for every set of realization rather than to minimize the \emph{average} number of samples.
Contrary to the work by Y{\i}lmaz \emph{et al.} \cite{yilmaz2015sequential,Yilmaz2014Sequential,Yilmaz2016Sequential}, we have investigated the problem of \gls{SJDE} in a series of works that aim to minimize the \emph{expected number of samples} while controlling the individual error probabilities and estimation error levels \cite{reinhard2022,reinhard2018,reinhard2024}.

\subsection{Aims and Outline of the Paper}
The aims of this work are as follows: First, we provide an introduction to \gls{JDE}.
We will limit our presentation to simple, easy-to-interpret performance measures and summarize the results published in various articles.
We will then illustrate the introduced concepts with a numerical example.
In particular, we show the performance gap to suboptimal two-step methods.
Unlike the first part, which assumes a fixed number of samples, the second part assumes a sequential setup.
Through sequential hypothesis testing, we introduce sequential analysis.
Next, the results from the \gls{fss} scenario are transferred to the sequential scenario.
A simple numerical example illustrates the properties of the optimal procedure and highlights its advantages over suboptimal sequential and optimal \gls{fss} procedures.
The third part discusses open challenges arising from the current results and new directions in (sequential) \gls{JDE} that should be addressed in the future.

The remainder of the paper is structured as follows:
\begin{itemize}
    \item In \textbf{\cref{sec:jde}}, the problem of \gls{JDE} using a fixed number of samples is investigated.
    First, the statistical model and the relevant performance measures are introduced.
    Next, widely used yet suboptimal procedures for the coupled detection and estimation problem are presented.
    Subsequently, the concept of optimal \gls{JDE} is  introduced, and the key properties are illustrated via a numerical example.
    Finally, extensions in terms of more complex performance measures that are used in literature are presented, and it is discussed how these affect the presented concepts.
    \item In \textbf{\cref{sec:sjde}}, the problem of \gls{JDE} is investigated in a sequential setup.
    First, the idea of sequential analysis is introduced, followed by a presentation of the sequential counterparts of the suboptimal methods presented in \cref{sec:subopt}.
    Then, the problem of optimal \gls{SJDE} is investigated, starting with a thorough problem formulation, followed by a step-by-step derivation of the optimal procedure and practical design algorithms that automatically chose the parameters on which the optimal solution depends.
    Finally, by means of a numerical example, we show the properties of optimal \gls{SJDE} procedures and highlight the differences over suboptimal sequential and optimal \gls{fss} procedures.
    \item In \textbf{\cref{sec:open_problems}}, open problems and future research directions are discussed.
    These include deepening existing theory, deriving advanced design algorithms and applying the idea of (sequential) joint detection and estimation to new areas.
    \item \textbf{\cref{sec:open_problems}} summarizes the paper and draws conclusions.
\end{itemize}

\subsection{Notations}
Upper and lower case variables indicate random variables and their realizations, respectively.
Normal font symbols and bold font symbols denote scalars and vectors/matrices.
For a vector $\bm a$, $\bm a^\top$ denotes the transpose of vector $\bm a$ and $\Tr(\bm A)$ denotes the trace of matrix $\bm A$.
We further use $\obsFSS$ as the shorthand notation for the collection $\obsSingle_1,\ldots,\obsSingle_N$.
The expectation and variance of a random variable $\RVsingle$ are written as $\E[\RVsingle]$ and $\Var[\RVsingle]$.
When integrals are taken over the entire domain of a quantity, the integration domain is dropped for the sake of compactness, e.g., $\E[\RVsingle] = \int_{-\infty}^\infty \obsSingle p(\obsSingle)\dInt\obsSingle \equiv \int \obsSingle p(\obsSingle)\dInt\obsSingle$.
The indicator function of an event $\mathcal{A}$ is denoted by $\ind{\mathcal{A}}$.
Finally, the dependency of functions, such as decision rules or estimators, on the data is often dropped to end-up with a compact notation. \section{Joint Detection and Estimation}\label{sec:jde}
In this section, the problem of coupled detection and estimation is discussed for the case when the number of samples is fixed.
First, the signal model and performance measures for the coupled detection and estimation problem are introduced in \cref{sec:sig_mod_perf_measures}.
Subsequently, suboptimal and optimal procedures are discussed in \cref{sec:subopt} and \cref{sec:opt}, respectively.

\subsection{Signal Model and Performance Measures}\label{sec:sig_mod_perf_measures}
The data can be generated under two different hypotheses, whereas under each hypothesis, the distribution of the data may depend on a set of parameters.
In the context of coupled detection and estimation, the parameter of interest is often modeled as a random variable itself, see, e.g. \cite{moustakides2012joint,jajamovich2012minimax,li2016optimal}, which yields the following model
\begin{align}\label{eq:model_jde}
\begin{split}
 \Hyp_0\!: & \; \RVfss \sim P(\obsFSS\given\Hyp_0, \param_0)\,,\; \paramRV_0\sim P(\param_0\given\Hyp_0)\,,\\
 \Hyp_1\!: & \; \RVfss \sim P(\obsFSS\given\Hyp_1, \param_1)\,,\; \paramRV_1\sim P(\param_1\given\Hyp_1)\,.
 \end{split}
\end{align}
Moreover, we assume that each hypothesis occurs with a known prior probability.
The reason for this assumption is that the upcoming presentation is more concise and easier to interpret.
If no prior probabilities are known, one can simply set $P(\Hyp_0)=P(\Hyp_1)=0.5$. 

Besides deciding between the two hypotheses, one is, depending on the decision, also interested in an accurate estimate of the underlying parameter.
Hence, the aim is to find a decision  of the form\footnote{To focus on concepts and avoid technical difficulties, we do not consider \emph{randomized} decision rules in this work. Under some mild conditions, such as that the probability distributions under both hypotheses do not impose any point masses, randomization is not required. In robust hypothesis testing, for example, least favorable densities often result in a test statistic that is not a continuous random variable and, hence, requires randomization \cite{fauss2021minimax,gul2016robust,gul2017minimax}.} $\decR[N](\obsFSS)\in\{0,1\}$ as well as two estimators $\est[N]{0}(\obsFSS)$ and $\est[N]{1}(\obsFSS)$.
As one is interested in the outcome of the hypothesis test as well as in the estimate.
One intuitive design criterion for a \emph{good} procedure would be to minimize the detection and estimation error levels simultaneously.
However, these are conflicting goals that cannot be achieved simultaneously.

Before suboptimal as well as optimal methods for the problem of coupled detection and estimation are discussed, the necessary performance measures have to be introduced.
As in classical hypothesis testing, the performance of the detector is measured by the Type-I and Type-II error probabilities, also referred to as probability of false alarm and probability of missed detection, which are defined as
\begin{align}\label{eq:def_det_err}
        \detErr[0] & = P(\decR[N]=1\given\Hyp_0)
        \qquad\text{and}\qquad
        \detErr[1] = P(\decR[N]=0\given\Hyp_1)\,.
\end{align}

The definition of a performance measure for the estimators is not as straightforward.
The reason for this is two-fold.
First, there exist a lot of different loss functions in (Bayesian) estimation, such as the squared-error loss or the absolute-error loss, to name just two.
Second, as detection and estimation are coupled, evaluating the estimation error also depends on the outcome of the decision rule.
In this work, the squared-error loss, a widely used loss function, is utilized and the performance of the estimators is quantified as
\begin{align}\label{eq:def_est_err}
        \estErr[0] & = \E\Bigl[\ind{\decR[N]=0}\big\lVert\est[N]{0}-\param_0\big\rVert^2\bgiven\Hyp_0\Bigr]\qquad\text{and}\qquad
        \estErr[1] = \E\Bigl[\ind{\decR[N]=1}\big\lVert\est[N]{1}-\param_1\big\rVert^2\bgiven\Hyp_1\Bigr]\,.
\end{align}
From \cref{eq:def_est_err}, one can see that this definition differs from the conventional \gls{MSE}, namely, that the estimation error is set to zero in case a wrong decision is made.
The reason for this is as follows.
Under different hypotheses, the parameters may belong to different physical quantities or even have different dimensionality.
Hence, in such situations, the calculation of an estimation error may be rendered meaningless or be even impossible.

To provide intuitive and easily interpretable results, the presentation in this work is restricted to the measures as defined in \cref{eq:def_est_err}.
In \cref{sec:diff_perf_measures}, the use of different performance measures and their impact on the results presented in \cref{sec:opt} are discussed.

\subsection{Suboptimal Procedures}\label{sec:subopt}
\begin{figure}
    \centering
 \includegraphics{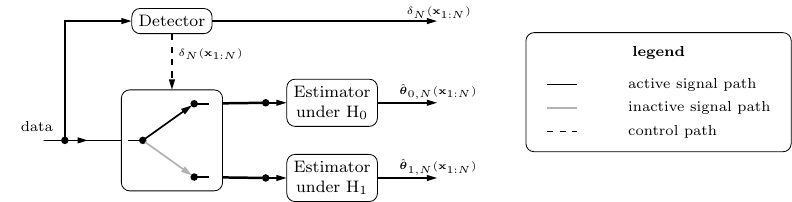}
 \caption{Suboptimal two-step procedure for a coupled detection/estimation problem. The detector computes the binary decision $\decR[N](\obsFSS)$, which controls the switch and routes the data to one of the two estimators. In the state shown, the switch routes the data to the estimator under $\Hyp_0$.}
 \label{fig:two-step}
\end{figure}
Before discussing optimal procedures for the coupled detection and estimation problem, some widely used procedures, yet not necessarily overall optimal ones are presented.
Often, such approaches are two-step methods as illustrated in \cref{fig:two-step}.
First, the data is passed to an optimal detector.
Then, it is passed to an estimator that is optimal under the chosen hypothesis.

One widely used optimality criterion in hypothesis testing is the one according to Neyman and Pearson.
Here, the idea is to design a detector that has a minimum Type-II error probability, while constraining the Type-I error probability to not exceed a nominal level $\detConstr[0]$.
Mathematically, this is equivalent to the optimization problem
\begin{align}\label{eq:opt_NP}
    \min_{\decR[N]}\;P(\decR[N]=0\given\Hyp_1)\,,\quad\text{s.t.}\quad P(\decR[N]=1\given\Hyp_0)\leq\detConstr[0]\,.
\end{align}
The optimal solution is given by \cite{neyman1933,poor2013introduction,levy2008principles,lehmann1986testing} 
\begin{align}\label{eq:decR_NP}
    \decR[N]^\text{NP}(\obsFSS) = \begin{cases}
                                        1 & \lr(\obsFSS) > \thNP\,, \\
                                        0 \;\text{{}or{}}\; 1 &  \lr(\obsFSS) = \thNP\,, \\
                                        0 & \lr(\obsFSS) < \thNP\,,
                                  \end{cases}
\end{align}
with the \gls{lr} given by
\begin{align*}
     \lr(\obsFSS) = \frac{p(\obsFSS\given\Hyp_1)}{p(\obsFSS\given\Hyp_0)}\,.
\end{align*}
In order for \cref{eq:decR_NP} to solve \cref{eq:opt_NP}, the threshold $\thNP$ has to be chosen such that the constraint in \cref{eq:opt_NP} is fulfilled with equality.
The resulting test is referred to as \gls{NP} test.

However, for composite hypotheses, the \gls{NP} test is not directly applicable.
One natural and widely used extension is the \gls{GLRT}, that is based on the idea of replacing the unknown parameter in the \gls{lr} by their maximum likelihood estimates, i.e., by the parameter that maximizes the conditional density function.
Mathematically, this can be written as
\begin{align}\label{eq:lr_glrt}
 \lrGLRT(\obsFSS) & = \frac{\max_{\param_1}p(\obsFSS\given\Hyp_1, \param_1)}{\max_{\param_0}p(\obsFSS\given\Hyp_0,\param_0)} = \frac{p(\obsFSS\given\Hyp_1, \hat\param_1^\text{ML})}{p(\obsFSS\given\Hyp_0, \hat\param_0^\text{ML})}\,,
\end{align}
where $\hat\param_0^\text{ML}$ and $\hat\param_1^\text{ML}$ denote the maximum likelihood estimates under the null hypothesis and the alternative, respectively.
After deriving the \gls{lr} as in \cref{eq:lr_glrt}, the \gls{NP} test can be constructed as outlined above.
There exist also other \gls{GLRT}-like detectors that replace the maximum-likelihood estimators in the likelihood ratio by estimators under different optimality conditions, such as the \gls{MAP} estimator.
It can be shown, that two-step procedures using \gls{GLRT}-like detectors admit optimality for certain formulations of the combined detection and estimation performance \cite{moustakides2009finite}.
However, this does not hold for general combined detection and estimation problems \cite{moustakides2012joint}.
Alternatively, as the model in \cref{eq:model_jde} assumes the parameter to be random, the likelihood ratio can be calculated as
\begin{align}\label{eq:lr_bayes}
 \lrBayes(\obsFSS) = \frac{\int p(\obsFSS\given\Hyp_1, \param_1) p(\param_1\given\Hyp_1)\dInt\param_1}{\int p(\obsFSS\given\Hyp_0, \param_0) p(\param_0\given\Hyp_0)\dInt\param_0}.
\end{align}
Similar as for the \gls{GLRT}, a Neyman-Pearson test can be designed using the likelihood ratio in \cref{eq:lr_bayes}.

Since the parameters are assumed to be random, we resort to a Bayesian optimality criterion for the estimator.
That is, the optimal estimator under hypothesis $\Hyp_i$, $i\in\{0,1\}$, can be found by solving
\begin{align*}
    \estOpt[N]{i}(\obsFSS) & = \argmin_a C_i(a\given\obsFSS) = \argmin_{a}\,\E[l(a, \param)\given\obsFSS]\,,\quad i\in\{0,1\}\,.
\end{align*}
In the equation above, $l(a,\param)\geq0$ denotes a loss function that assigns a particular cost to a pair of action (the estimate) $a$ and the parameter $\param$ and $C_i(a\given\obsFSS)$ is referred to as posterior loss.
As mentioned earlier in this section, this work uses the \gls{MSE} to quantify the performance of the estimator.
Hence, with a squared-error loss function, the optimal Bayes estimator becomes the posterior mean
\begin{align}\label{eq:post_mean}
    \estOpt[N]{i}(\obsFSS) = \E[\paramRV_i\given\Hyp_i,\obsFSS]
\end{align}
and the posterior loss, which measures the quality of the estimate, results in the posterior variance
\begin{align*}
    C_i(\est[N]{i}(\obsFSS)\given\obsFSS) = \Var[\paramRV_i\given\Hyp_i,\obsFSS]\,.
\end{align*}
For more details about estimation theory, see, for example, \cite{poor2013introduction,levy2008principles,lehmann1998theory}.

Now, as the optimal detector as well as the optimal estimators have been derived, a two-step procedure as depicted in \cref{fig:two-step} could be assembled.
However, as stated in \cite{moustakides2012joint}, even though the detector and the estimators are optimal themselves, the combined procedure is not necessarily overall optimal.
The difference to optimal procedures will be illustrated in \cref{sec:example_fss} by means of a numerical example.

\subsection{Optimal Procedures}\label{sec:opt}
In what follows, we discuss how the problem of coupled detection and estimation can be addressed in an optimal manner.
The main idea of optimal joint detection and estimation is to control both, the detection and the estimation performances.
As both performance measures cannot be minimized simultaneously, the power of the detector can be slightly reduced to give room for improving the quality of the estimates.

There exist two branches\footnote{Although this work distinguishes between the purely Bayesian and the \gls{NP}-like formulation, there also exist combination of both, such as constraining the overall error probability instead of the individual error probabilities \cite[Eq. (P2)]{li2016optimal}.} of how the problem of coupled detection and estimation is investigated in a jointly optimal manner: a purely Bayesian formulation such as in \cite{middleton1968simultaneous,fredriksen1972simultaneous,li2007optimal,milder2014simultaneous} and a \gls{NP}-like formulation such as in \cite{moustakides2012joint,jajamovich2012minimax,tajer2010optimal,li2016optimal}.

Under the purely Bayesian framework, one introduces a cost function that combines the detection and estimation error levels.
Subsequently, a decision rule and a set of estimators are to be found that minimize the cost function.
Mathematically, this problem can be written as
\begin{align}\tag{P1}\label{eq:jde_prob_bayes}
 & \min_{\decR[N], \est[N]{0}, \est[N]{1}} J^\text{B}(\decR[N], \est[N]{0}, \est[N]{1})\,.
\end{align}
Here, the trade-off between detection and estimation performance is not given explicitly in the optimization problem, but given through the definition of the objective function.
Although the exact form of this objective function is highly problem dependent, it is usually some kind of linear-combination of detection and estimation error levels.
This kind of problem formulation is common in a decision theoretic framework, however, it comes with a major drawback for the combined detection and estimation problem: how to choose the weight coefficients for the detection and estimation performance measures.
One reason is that detection and estimation error levels usually have a different numerical range.
Additionally, the error probabilities decay with $\mathcal{O}(e^{-N})$, whereas the estimation errors decay only with $\mathcal{O}(N^{-1})$.

Under the \gls{NP}-like framework, one aims at minimizing the estimation error levels under constraints on the error probabilities.
More formally, this problem can be written as
\begin{align}\tag{P2}\label{eq:jde_prob_np}
    \begin{split}
 & \min_{\decR[N], \est[N]{0}, \est[N]{1}} J^\text{NP}(\decR[N], \est[N]{0}, \est[N]{1})\,, \\
\text{s.t.}  \quad & P(\decR[N]=1\given\Hyp_0)\leq\detConstr[0]\,, \quad P(\decR[N]=0\given\Hyp_1) \leq \detConstr[1]\,,
 \end{split}
\end{align}
with the nominal error probabilities $\detConstr[0],\detConstr[1]\in(0,1)$.
For \cref{eq:jde_prob_np} to admit a solution, the nominal Type-II error level must not be smaller than the Type-II error probability of an \gls{NP} test with Type-I error probability $\detConstr[0]$, i.e, $\detConstr[1]\geq \detErr[1]_\text{NP}(\detConstr[0])$.
As the outcome of the decision determines which estimator is applied, the cost function quantifying the estimation error levels, depends on the decision rule as well as the estimators.
The main idea of this problem formulation is to relax the requirements on the detector, i.e., allowing a higher Type-II error probability than the \gls{NP} test, and using this space to improve the performance of the estimator in the coupled case.

In what follows, the solutions for both problem formulations are discussed under the assumption that simple objective functions are used
\begin{align}
    J^\text{B}(\decR[N], \est[N]{0}, \est[N]{1}) & = \sum_{i=0}^1 p(\Hyp_i)\Bigl(\detCost\detErr + \estCost\estErr\Bigr)\,,\label{eq:J_B}\\
    J^\text{NP}(\decR[N], \est[N]{0}, \est[N]{1}) & = \sum_{i=0}^1 p(\Hyp_i)\estCost\estErr\,.\label{eq:J_NP}
\end{align}
Here, the objective in the Bayesian case is simply a linear combination of all performance measures, whereas it is a linear combination of the estimation error levels in the \gls{NP} case.
For both objective functions and with the performance measures defined in \cref{eq:def_det_err} and \cref{eq:def_est_err}, respectively, it can be shown that the optimal estimator is independent of the decision rule.
When using the optimal estimator in \cref{eq:post_mean}, the cost for accepting hypothesis $\Hyp_i$, $i\in\{0,1\}$, is given by
\begin{align}\label{eq:def_D}
    \begin{split}
    D_i(\obsFSS) & =  \detCost[1-i]p(\Hyp_{1-i}\given\obsFSS) + \estCost[i]p(\Hyp_i\given\obsFSS)\Var[\paramRV_i\given\Hyp_i,\obsFSS]\,,
    \end{split}
\end{align}
when the optimal Bayes estimator from \cref{eq:post_mean}, i.e., the posterior mean, is used.
The cost function in \cref{eq:def_D} composes of two parts, the first one penalizes a wrong decision, whereas the second one penalizes inaccurate estimates.
Hence, one can see that the cost for accepting $\Hyp_i$, $i\in\{0,1\}$, depends on both, the confidence about the hypothesis and the parameter under which the data was generated. The optimal decision rule can then be written as
\begin{align}\label{eq:def_dec_fss}
    \decROpt[N](\obsFSS) & = \begin{cases}
                                1 & D_0(\obsFSS) > D_1(\obsFSS)\,, \\
                                0 \;\text{{}or{}}\; 1 & D_0(\obsFSS) = D_1(\obsFSS)\,, \\
                                0 & D_0(\obsFSS) < D_1(\obsFSS)\,,
                              \end{cases}
\end{align}
and the overall cost is denoted by
\begin{align}\label{eq:def_g}
    g(\obsFSS) = \min\{D_0(\obsFSS)\,,\,D_1(\obsFSS)\}\,.
\end{align}
Though it might look like that estimators are independent of the decision rule, the estimator $\estOpt[N]{i}$, $i\in\{0,1\}$, is triggered if and only if a decision in favor of $\Hyp_i$, $i\in\{0,1\}$, is made.
This means that the estimators and the decision rule are fully coupled.

From \cref{eq:def_D}, one can see that the cost for accepting hypothesis $\Hyp_i$ and, as a consequence of this, also the detection and estimation performances, depend on the coefficients $\detCost$, $\estCost$, $i\in\{0,1\}$.
When using the Bayesian formulation, these coefficients are set by the practitioner and their choice is highly application dependent.
Under the \gls{NP}-like formulation, the coefficients $\estCost$, $i\in\{0,1\}$, are still to be set by the practitioner.
These coefficients allow balancing both estimation error levels that are, for example, caused by different numerical ranges of the parameters.
The coefficients $\detCost$, $i\in\{0,1\}$, have to be chosen such that the constraints in \cref{eq:jde_prob_np} are fulfilled with equality.
Following the ideas stated, e.g., in \cite{fauss2015linear,reinhard2018}, it can be shown that these coefficients act as Lagrange multipliers and can be found by solving the convex optimization problem
\begin{align}\label{eq:opt_coeff_fss}
    \max_{\detCost[0]\geq0, \detCost[1]\geq0}\; \E\Bigl[g(\obsFSS)\Bigr] - \sum_{i=0}^1 p(\Hyp_i)\detCost\detConstr\,.
\end{align}
As long as the problem is feasible, i.e., $\detConstr[1]\geq\detErr[1]_\text{NP}$, the decision rule parametrized by the solution of \cref{eq:opt_coeff_fss} also solves \cref{eq:jde_prob_np}.

\subsubsection{Illustrative Example}\label{sec:example_fss}
To provide more insight into the different problem formulations as well as their optimal solutions, a numerical example is presented.
Here, we resort to a more simplistic model than \cref{eq:model_jde}, namely, the case where a parameter of interest is only present under the alternative.
This model is simple enough to show the properties of the formulation and optimal solution while still covering important applications such as the simultaneous detection and localization of targets in radar.
Mathematically, the model reduces to
\begin{align}
    \begin{split}
        \Hyp_0\!: & \; \RVfss \sim P(\obsFSS\given\Hyp_0)\,,\\
        \Hyp_1\!: & \; \RVfss \sim P(\obsFSS\given\Hyp_1, \param_1)\,,\; \paramRV_1\sim P(\param_1\given\Hyp_1)\,.
    \end{split} \label{eq:model_simple}
\end{align}
Joint detection and estimation for this model has first been investigated by Moustakides \emph{et al.} \cite{moustakides2012joint} resulting in a one-step test and a two-step test.
For the one-step test, it can be shown that it admits a similar structure as the decision rule in \cref{eq:def_dec_fss}.
Contrary to this, the aim of the two-step test is not to reduce the detection power to decrease the estimation error levels, but only provide an estimate if it seems to be accurate.
Therefore, a second test stage is introduced that decided if an estimate should be provided or not.
However, this is beyond the scope of this work and the interested reader is referred to \cite[Sec. II-B]{moustakides2012joint}.

We assume that under both hypotheses, the data conditioned on the mean follows a Gaussian distribution.
Under the null hypothesis, the mean is constant and set to $\theta_0=0$, whereas the mean follows a uniform distribution on $[1,6)$ under the alternative.
The variance is set to $\sigma^2=9$ and $N=10$ samples are observed.
As only the model under the alternative involves a parameter of interest, we set $\estCost[1]=1$,

To illustrate the different decision rules, we use the sample mean $\bar{x}_N$  as sufficient statistic or state of the procedure as it captures all relevant information contained in $x_1,\ldots,x_N$.

The two-step procedure outlined in \cref{sec:subopt}, which comprises an \gls{NP} test and a \gls{mmse} estimator, is used for benchmarking.
When setting the nominal probability of false alarm to $5\,\%$, the \gls{NP} test  achieves a probability of missed detection of approximately $14.7\,\%$.

To illustrate the properties of the optimal \gls{JDE} procedure under the \gls{NP}-like formulation, we fix the probability of false alarm to $5\,\%$ and restrict the probability of missed detection to different values $\detConstr[1] = \detErr[1]_\text{NP} + \Delta\detErr[1]$.
More precisely, we sweep the relaxation of the Type-II error $\Delta\detErr[1]$, from $\Delta\detErr[1]=0\,\%$, which is equivalent to the \gls{NP} test, to $\Delta\detErr[1]=35\,\%$.
\begin{figure}
\subfloat[ Empirical error probabilities and empirical \gls{MSE} for different values of $\Delta\alpha^1$. The results for the \gls{NP} test are marked with a circle ($\circ$). Target error probabilities are shown as a dashed line (\dashed). \label{fig:jde_err}]{
        \includegraphics[]{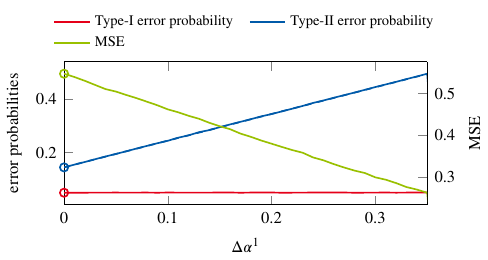}
    }\hfill
    \subfloat[Decision rule for different values of the relaxation of the Type-II error $\Delta\alpha^1$. No relaxation (\textcolor{my_green}{$\bm -$}), i.e., $\Delta\alpha^1=0$, corresponds to the \gls{NP} test. \label{fig:jde_dec}]{
        \includegraphics[]{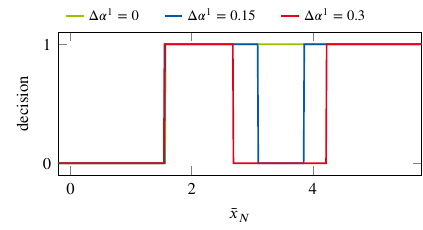}
    }
    \caption{Comparison of different \gls{NP}-like optimal procedures for different relaxations of the Type-II error probabilities $\Delta\alpha^1\in\{0,0.15,0.3\}$. The  optimal coefficients are given by $(\detCost[0],\detCost[1])\in\{ (0.51, 0.84), (0.48, 0.82), (0.40, 0.77) \}$.}
    \label{fig:jde}
\end{figure}
In \cref{fig:jde_err}, the empirical error measures for different target values $\detConstr[1] = \detErr[1]_\text{NP} + \Delta\detErr[1]$ are shown.
It can be seen that the empirical error probabilities hit the targeted ones exactly.
Additionally, it can be seen as the targeted Type-II increases linearly, the empirical \gls{MSE} reduces linearly.
This coincides with the main idea of \gls{JDE} as relaxing the Type-II error probability should result in a procedure with reduced \gls{MSE}.
How this effect is achieved can be seen in \cref{fig:jde_dec}, which  shows the decision rule for different values of $\Delta\detErr[1]$.
It can be seen that the \gls{NP} test ($\Delta\detErr[1]=0$) decides in favor of the alternative once the sample mean is larger than a threshold ( $\approx1.57$). 
As a high sample mean results in a high confidence that the alternative is true, this result is consistent with what one would expect.
If one increases $\Delta\detErr[1]$, the form of the decision rule changes drastically.
Whereas the above-mentioned threshold remains, a corridor centered around $\bar{x}_N=3.5$ appears that is widened when $\Delta\detErr[1]$ increases.
Though the certainty that the alternative is true is high in these regions, the uncertainty about the true parameter is also high.
Hence, by accepting more incorrect decisions, the overall estimation error levels can be decreased.

\begin{figure}
    \centering
    \includegraphics[]{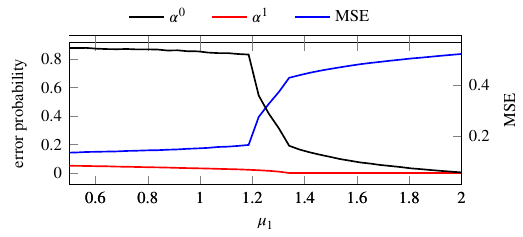}
    \caption{Performance measures for different values of $\estCost[1]$ in the Bayesian formulation. The cost for making a wrong decision are set to $\detCost[0]=\detCost[1]=1$. }
    \label{fig:jde_Bayes}
\end{figure}

Besides the \gls{NP}-like formulation, there also exist a Bayesian formulation in which costs are assigned to every kind of error.
This approach is well known in Bayesian hypothesis testing.
However, in the context of \gls{JDE} setting these costs by hand is non-trivial.
To illustrate this, we fixed the cost for making a wrong decision to $\detCost[0]=\detCost[1]=1$ and swept the cost for an inaccurate estimate.
The results are shown in \cref{fig:jde_Bayes}.
Here, one can see that the performance measures are highly non-linear and non-smooth in $\estCost[1]$, which makes a choice by the practitioner highly challenging.

\subsection{Different Estimation Performance Measures}\label{sec:diff_perf_measures}
So far, we have quantified the overall estimation errors as a linear combination of the individual \glspl{MSE} to provide a simple and conceptual presentation.
However, a meaningful quantification of the estimation performance may highly depend on the application and on the underlying signal model.
In this section, we present a different formulation of the error metrics that arise in the \gls{JDE} literature and discuss how these will affect the results presented above.

In \cite{moustakides2012joint,jajamovich2012minimax}, for example, the estimation are not simply set to zero in case of a wrong decision.
Instead, the measure is conditioned on a correct decision, i.e.,
\begin{align*}
    \estErr & = \E\Bigl[l(\est[N]{i},\paramRV_i)\bgiven\Hyp_i,\decR[N]=i\Bigr] = \frac{\E\Bigl[\ind{\decR[N]=i}l(\est[N]{i},\paramRV_i)\bgiven\Hyp_i\Bigr]}{P(\decR[N]=i\given\Hyp_i)}\,,
\end{align*}
which can be seen to be a non-linear function of the decision  rule $\decR[N]$.
The linearity of all performance measures in the decision rule is an important property in the derivation of the optimal decision rule and the calculation of the optimal cost coefficients as described earlier.

Additionally, a second non-linearity in the decision rule is introduced in \cite{jajamovich2012minimax}.
Instead of using a linear combination of the individual estimation error levels to quantify the overall estimation performance as used in \cref{eq:J_NP}, the maximum of the estimation error levels is used in \cite{jajamovich2012minimax}.
That is, the objective to be minimized is of the form
\begin{align*}
    J(\decR[N], \est[N]{0},\est[N]{1}) = \max_{i\in\{0,1\}} \estErr\,.
\end{align*}
Although the decision rule presented in \cite{jajamovich2012minimax} shows structural similarities to the one presented above in the sense that it involves the likelihoods under both hypotheses and the posterior losses, it depends on four parameters.
The optimal parameters have then to be obtained via grid search \cite[Table 1]{jajamovich2012minimax}.

Another important, application-dependent question is how to characterize estimation errors in the case of an incorrect decision.
Contrary to the present work, \cite{tajer2010optimal, Yilmaz2016Sequential,Shang2015,li2016optimal} distinguish between the estimation errors in the case of a correct or an incorrect decision.
That is, the estimation error under $\Hyp_i$, $i\in\{0,1\}$, is quantified as
\begin{align*}
    \estErr = &b_{i0}\E\Bigl[\ind{\decR[N]=0}l(\est[N]{0},\paramRV_0)\bgiven\Hyp_i\Bigr] +  b_{i1}\E\Bigl[\ind{\decR[N]=1}l(\est[N]{1},\paramRV_1)\bgiven\Hyp_i\Bigr]\,,
\end{align*}
where $b_{i0}, b_{i1}$, $i\in\{0,1\}$, are some non-negative cost coefficients.
As a result, the optimal estimators are a linear combination of the optimal Bayes estimators under both hypotheses.
Although this slightly affects the exact form of the optimal decision rule, the main results stated so far are still valid.

If the parameters are vector-valued, there are more aspects that may be considered when defining the estimation performance.

Depending on the application, the individual parameters under a particular hypothesis may be of different numerical range.
In such situations, the \gls{MSE} as defined in \cref{eq:def_est_err} might not be beneficial as the overall \gls{MSE} is dominated by one parameter.
Therefore, introducing a weighted \gls{MSE} of the form
\begin{align*}
\estErr = \E\Bigl[\ind{\decR[N]=i}\!\Tr\Bigl((\est[N]{i}-\param_i)\mathbf{W}(\est[N]{i}-\param_i)^\top\Bigr)\bgiven\Hyp_i\Bigr]
\end{align*}
for some weight matrix $\mathbf{W}$, might be reasonable to account for the different numerical range.
The diagonal elements of $\mathbf{W}$ act as weights for the individual variances.
Although this does not affect the optimal estimators, one can tune how the estimation errors of individual parameters contribute to the decision rule.
Nevertheless, the general results that have been presented in this section are still valid.

Consider the toy example of testing a zero-mean Gaussian distribution against a Gaussian distribution with positive and random mean.
Under both hypotheses the variance is assumed to be random.
In addition to hypothesis testing, we wish to estimate the parameters of the chosen model.
Here, the variance is a parameter of interest under both hypotheses and the mean is only of interest if we decide in favor of the alternative.
Therefore, it might be beneficial to proceed as, for example, in \cite{Dulek2018,li2016optimal}, and to split each parameter vector into two sets of parameters.
The first one contains the parameters that are used in the models under both hypotheses, whereas the second set contains the remaining parameters.
This formulation has the advantage of a flexible yet easy to interpret way to define the objective quantifying the estimation quality.
The resulting optimal estimator highly depends on the exact formulation, but the principle analysis that has been presented in this section is still valid.

One can conclude that there is a variety of ways to quantify the quality of the estimators.
The particular choice depends on the problem at hand.
However, the practitioner should keep in mind that the choice of the performance measure for the estimators may also affect the complexity of the design process.

 \section{Sequential Joint Detection and Estimation}\label{sec:sjde}
This section discusses the problem of \gls{JDE} when the observations arrive as streaming data and one wishes to use on average as few samples as possible while imposing constraints on the inference quality.
First, an introduction to sequential analysis is given.
Subsequently, suboptimal and optimal sequential procedures for \gls{JDE} are presented.

\subsection{Fundamentals of Sequential Analysis}
Contrary to the methods addressed before, the number of samples is not given a priori for sequential methods.
Instead, the observations arrive sequentially and one stops collecting new samples once the observed ones are significant enough.
This idea goes back to the late 1940s when Wald invented the \gls{sprt} \cite{wald1945sequential,wald1948}.

To decide whether to stop or to continue sampling, any sequential procedure is equipped with a stopping rule of the form \footnote{Similar to the decision rule in the previous section, we assume that no randomization is necessary. However, for robust sequential procedures a randomized stopping rule is most probably required \cite{fauss2020minimax,fauss2021minimax}.} $\stopR\in\{0,1\}$,
which results in the definition of the run-length or stopping time
\begin{align}\label{eq:seq_rl}
\rl = \min\{n\geq0: \stopR=1\}\,.
\end{align}
Note that the stopping rule evaluates the observations and, hence, the outcome of the stopping rule and the run-length are random variables.
Finding a stopping rule such that one stops sampling after a certain level of confidence is reached is the core of sequential procedures.

\begin{figure}
     \centering
\includegraphics{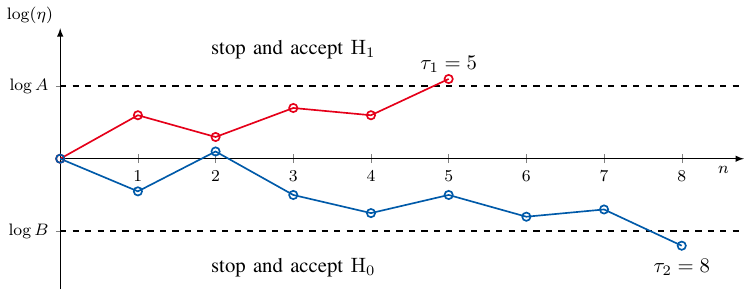}
    \caption{Example of a Sequential Probability Ratio Test.}
    \label{fig:sprt_example}
\end{figure}

To illustrate the concept of sequential analysis, the problem of sequential hypothesis testing is considered.
In sequential hypothesis testing, the goal is to find a procedure consisting of a stopping rule and a decision rule that minimizes the \emph{expected} number of samples while keeping the error probabilities below the nominal error levels.
Contrary to the \gls{fss} scenario, the number of samples as additional degree of freedom allows the control both error probabilities.
More formally, the design problem reads
\begin{align}
     \min_{\{\stopR,\decR\}_{n\geq0}}\;\E[\tau]\,,\quad\text{s.t.}\quad\detErr[0]\leq\detConstr[0]\,,\quad\detErr[1]\leq\detConstr[1]\,.
\end{align}
This results in a \gls{lr} test that updates its test statistic when a new sample is observed rather than calculating the test statistic for all samples at once.
Wald proposed a test with two thresholds $0<B<A$ that stops and decides in favor of the null hypothesis/the alternative once the likelihood ratio crosses the lower/upper threshold and continues sampling otherwise \cite{wald1945sequential,wald1948}.
A test of this form is illustrated in \cref{fig:sprt_example}.
More formally, the stopping and decision rule can be expressed as
\begin{align}
     \stopR(\obsSeq) = \begin{cases}
          0 & \lr(\obsSeq) \in (A,B)\,, \\
          1 & \text{else}\,,
     \end{cases}
\end{align}
and
\begin{align}
     \decR(\obsSeq) = \begin{cases}
          0 & \lr(\obsSeq) \leq B\,, \\
          1 & \lr(\obsSeq) \geq A\,.
     \end{cases}
\end{align}
The choice of the strictly optimal thresholds is non-trivial.
However, Wald suggested \cite{wald1945sequential} to approximate the threshold as
\begin{align}\label{eq:wald_th}
     \begin{split}
          A & = \frac{1-\detConstr[1]}{\detConstr[0]}\,
          \quad\text{and}\quad
          B = \frac{\detConstr[1]}{1-\detConstr[0]}\,,
     \end{split}
\end{align}
which have shown to become optimal as the nominal error levels tend to zero \cite{wald1948}.
For non-zero nominal error levels, however, the test results in much smaller error probabilities than the nominal ones, which leads to an increase in the expected run-length.
Contrary to this, there exist optimal sequential tests \cite{novikov2009optimal,novikov2009multiple,fauss2015linear}, i.e., test that hit the targeted error probabilities exactly while strictly minimizing the expected number of used samples.
Strictly optimal sequential procedures will become of particular importance in \cref{sec:opt_seq_jde}.

Since the invention of the \gls{sprt}, design and analysis of sequential detectors and sequential estimators has become important.
An overview of sequential detection and sequential estimators is, for example, given in \cite{tartakovsky2014sequential} and \cite{ghosh2011sequential}, respectively.

\subsection{Suboptimal Sequential Procedures}
Next, we briefly discuss suboptimal sequential procedures for the problem of coupled detection and estimation.
In principle, the procedures presented in \cref{sec:subopt} can directly be translated to the sequential scenario.
Before discussing technical details, the structure of the two-step procedure needs close inspection.
In the sequential case, the two-step as depicted in \cref{fig:two-step} comprises a sequential detector followed by a \gls{fss} estimator.
This detector determines, entirely based on the certainty about the data generating hypothesis, whether to take an additional sample or not.
Once the decision to stop sampling is made, all samples collected so far are passed to the estimator that has to infer the unknown parameter(s).
In the sequential case, the suboptimality of the two-step procedure can be seen directly as the uncertainty about the unknown parameter(s) is not considered when the number of samples is determined.

One way to realize a sequential detector for composite hypotheses is to design it based on some generalized likelihood ratio.
Here, all available samples may be used to estimate the unknown parameters, which results in the \gls{gsprt} \cite[Section 5.4.1]{tartakovsky2014sequential}. When not all available samples are used to estimate the unknown parameters, the resulting test is often referred to as adaptive \gls{sprt} \cite[Section 5.4.2]{tartakovsky2014sequential}.
At this point, it should be mentioned that when using some generalized test statistic, the resulting test might not fulfill the constraints on the error probabilities because the estimator introduces additional uncertainty that is not considered in the stopping rule.
Hence, especially for small sample sizes the sequential test might stop and make an erroneous decision.
To overcome this, the bootstrap can be used to account for the additional uncertainty introduced by the estimator \cite{golz2017bootstrapped}.
In a Bayesian framework, one can use the likelihood ratio of the form \cref{eq:lr_bayes}, which will be used as a reference in this work.

\subsection{Optimal Sequential Procedures}\label{sec:opt_seq_jde}
The problem of joint detection and estimation has first been investigated in a sequential setup in the late 2000s \cite{buzzi2006,grossi2008,grossi2009sequential}.
In that works, the outcome of the detector and of the estimator are of primary interest, however, the stopping rule does not take parameter uncertainty into account.
Y{\i}lmaz \emph{et al.} investigated the problem of \gls{SJDE} in \cite{Yilmaz2014Sequential,yilmaz2015sequential,Yilmaz2016Sequential}.
However, these works do not aim to minimize the \emph{average} number of used samples, but rather the number of observations for every run.
Though this assumption significantly decreases the design complexity, it is not in line with the classical formulation in sequential inference and increases the average number of used samples.
Besides these works, we have investigated the problem of \gls{SJDE} in a series of publications \cite{reinhard2016approach,fauss2017sequential,reinhard2018,reinhard2019a,reinhard2020a,reinhard2020b,reinhard2021,reinhard2021distributed,reinhard2022,reinhard2024}.
In that works, the aim is to minimize the \emph{expected} number of used samples while the error probabilities and the estimation error levels are kept below nominal levels.

Next, optimal solutions to the problem of sequential joint detection are presented, which are mainly based on the work in \cite{reinhard2018}.
For the sake of simplicity, only the case of binary hypothesis testing is considered.
An extension to the $M$-ary scenario can be found in \cite{reinhard2022,reinhard2020a}.

First, let $\mathcal{X}=(\RVsingle_n)_{n\geq1}$ denote a sequence of random variables that can be generated under one out of two hypotheses and, under each hypothesis, the data generating probability measure depends on a random parameter with known distribution.
Mathematically, the model for the problem of joint detection and estimation reads as
\begin{align}\label{eq:model_sjde}
\begin{split}
 \Hyp_0\!: & \; \mathcal{X} \sim P(\obsSingle_1,\obsSingle_2, \ldots \given\Hyp_0, \param_0)\,,\; \paramRV_0\sim P(\param_0\given\Hyp_0)\,,\\
 \Hyp_1\!: & \; \mathcal{X} \sim P(\obsSingle_1,\obsSingle_2, \ldots \given\Hyp_1, \param_1)\,,\; \paramRV_1\sim P(\param_1\given\Hyp_1)\,.
 \end{split}
\end{align}

Besides the decision rule and the two estimators as in the \gls{fss} scenario in \cref{sec:jde}, the sequential procedure is also equipped with a stopping rule.
The collection of stopping rule, decision rule and the two estimators is referred to as policy from now on and is defined as $\policy:=\{\stopR\,, \decR\,, \est{0}\,, \est{1}\}_{n\geq0}$. 
The aim is to find a policy that minimizes the \emph{expected} number of used samples while the detection and estimation error levels are to be kept below nominal ones.
Mathematically, this can be written as the optimization problem
\begin{align}\label{eq:problem_sjde}
     \begin{split}
          \min_\policy\;\E[\tau]\,,\;\;\text{s.t.}\;\;&\detErr[0]\leq\detConstr[0]\,,\;\detErr[1]\leq\detConstr[1]\,,\\
          & \estErr[0]\leq\estConstr[0]\,,\;\estErr[1]\leq\estConstr[1]\,.
     \end{split}
\end{align}
Assuming that the stopping time $\tau$ is fixed, the optimal decision rule and the optimal estimators are equal to those for the \gls{fss}, see \cref{eq:def_dec_fss} and \cref{eq:post_mean}.
Recall from the \gls{fss} scenario presented in \cref{sec:opt}, that the overall cost after $\obsSeq$ have been observed is given by
\begin{align}\label{eq:def_g_seq}
     g(\obsSeq) = \min\bigl\{D_0(\obsSeq)\,,\,D_1(\obsSeq)\bigr\}\,,
\end{align}
with the cost for accepting $\Hyp_i$, $i\in\{0,1\}$,
\begin{align}\label{eq:def_D_seq}
     D_i(\obsSeq) & =  \detCost[1-i]p(\Hyp_{1-i}\given\obsSeq) + \estCost[i]p(\Hyp_i\given\obsSeq)\Var[\paramRV_i\given\Hyp_i,\obsSeq]\,.
\end{align}
Again, it has to be highlighted that the cost for accepting $\Hyp_i$, $i\in\{0,1\}$, depends on two terms: one for penalizing a wrong decision and one for penalizing inaccurate estimates.
The cost coefficients $\detCost, \estCost$, $i\in\{0,1\}$, are assumed to be fixed for now.
Their choice will be discussed later in this section.
In the sequential scenario, the function defined in \cref{eq:def_g_seq} describes the cost after observing $\obsSeq$.
Now, the aim is to find an optimal stopping rule that minimizes the overall cost.
Determining an optimal stopping rule is the most challenging part when designing optimal procedures.
To simplify notation, we introduce the shorthand notation
\begin{align*}
     \stopAt(\obsSeq) = \stopR(\obsSeq)\prod_{i=0}^{n-1}\Bigl(1-\stopR[i](\obsSeq[i])\Bigr)
\end{align*}
for a function indicating that the procedure stops sampling at time $n$ and not before.
More precisely, $\stopAt(\obsSeq)=1$ if we stop sampling at time $n$ given observations $\obsSeq$ and we have not stopped before and  $\stopAt(\obsSeq)=0$ else.
Finally, the problem of finding the optimal stopping rule is given by the following optimization problem
\begin{align}
     \min_{\{\stopR\}_{n\geq0}}\;\sum_{n=0}^\infty\E\Bigl[\stopAt(\obsSeq)(n+g(\obsSeq))\Bigr]\,.
\end{align}
Solving problems of this type, which are also referred to as \emph{optimal stopping problems}, can be tedious. A general treatment of these problems is beyond the scope of this work.
For an overview of fundamental concepts in optimal stopping theory see, for example, \cite{Peskir06,Chow71Theory,poor2009quickest}.

Before the optimal stopping rule can be derived, a sufficient statistic $\stat$, that acts as the state of the sequential procedure, has to be introduced.
This low-dimensional representation $\stat$ captures all statistical information contained in $\obsSeq$ about the underlying hypotheses $\Hyp_i$, $i\in\{0,1\}$, the random parameters $\paramRV_i$, $i\in\{0,1\}$, and the next sample $\RVsingle_{n+1}$.
The initial state is defined as $\stat[0]$.

When the maximum number of used samples is restricted to not exceed $\Nmax$, the optimal procedure can be characterized by the non-linear Bellman equation
\begin{align}\label{eq:def_rho}
\begin{split}
          \rho_n(\stat) &= \min\Bigl\{g(\stat)\,,\,\contCost(\stat)\Bigr\}\,,\;n<\Nmax\,, \\
          \rho_{\Nmax}(\stat[\Nmax]) &= g(\stat[\Nmax])\,.
     \end{split}
\end{align}
Here, $\rho_n(\stat)$ defines the overall cost function, $g(\stat)$ as defined in \cref{eq:def_g} the cost for stopping sampling and
\begin{align}
\contCost(\stat) = 1+\E\bigl[\rho_{n+1}(\stat[n+1])\given\stat\bigr]\,,\;n<\Nmax\,,
\end{align}
is the cost for taking an additional sample.
Given state $\stat$ at time $n$, one chooses the action, i.e., stop sampling or take an additional sample, that has minimum cost.
Hence, one stops sampling as soon as $g(\stat)<d_n(\stat)$, which leads to the optimal stopping rule:
\begin{align*}
     \stopRopt(\stat) = \begin{cases}
          1 & g(\stat) < \contCost(\stat)\\
          0 \;\text{{}or{}}\; 1 & g(\stat) = \contCost(\stat) \\
          0 & g(\stat) > \contCost(\stat)
     \end{cases}
\end{align*}
To evaluate the stopping rule, the state space of the sufficient statistic has first to be discretized, and the cost functions have to be calculated recursively based on the definition in \cref{eq:def_rho}.
This significantly complicates the design when the dimensionality of $\stat$ and/or the maximum number of samples $\Nmax$ increases.
Some aspects how this can be overcome will be addressed in the remainder of the paper.

Though the procedure presented so far is optimal, the policy still depends on $\detCost, \estCost$, $i\in\{0,1\}$, and, hence, these coefficients have to be chosen such that the policy solves \cref{eq:problem_sjde}.
It can be shown that the optimal coefficients are given by the solution of the following convex optimization problem \cite{reinhard2018}
\begin{align}\label{eq:seq_opt_coeff}
     \max_{\detCost, \estCost \geq 0}\;\rho_0(\stat[0]) - \sum_{i=0}^1p(\Hyp_i)\Bigl(\detCost\detConstr + \estCost\estConstr\Bigr)\,.
\end{align}
Assuming that the dimensionality of $\stat$ is small, the optimization problem can be cast into a \gls{lp} \cite{reinhard2018,fauss2015linear} that can efficiently be solved by powerful off-the-shelf solvers.
Besides that, gradient based techniques, such as a projected gradient ascent \cite{reinhard2019a} or a \gls{bfgs} method \cite{reinhard2024} have been used to solve this problem.

Due to the recursive definition of the cost function that defines the stopping rule, the design and implementation of optimal sequential procedures can become very demanding.
To overcome this, an \gls{apo} procedure \cite{reinhard2021} and an \gls{ao} procedure \cite{reinhard2024} were introduced.
The main idea behind these procedures is to provide an easy-to-design and easy-to-implement procedure whose performance is close to optimal and even becomes optimal as the nominal error levels tend to zero.
Under some additional assumptions, it has been shown that the stopping rule
\begin{align}
     \stopRao(\obsSeq) = \begin{cases}
               1 & \frac{g(\obsSeq)}{n+1} \leq 1 \\
               0 & \text{else}
     \end{cases}
\end{align}
is \gls{ao} \cite{reinhard2024}.
Contrary to the optimal stopping rule, it does not require the calculation of a recursive cost function, but can instead be evaluated on the fly.
Furthermore, it does not even require the existence of a sufficient statistic.
The coefficients that parameterize the \gls{ao} policy can be determined by a \gls{bfgs} method \cite{reinhard2024}.

However, one drawback of the \gls{ao} stopping rule is that a parameter to be estimated has to exist under all hypotheses.
Hence, it is not applicable to simplified models as, e.g., \cref{eq:model_simple}.

\subsection{Numerical Example}
The model from \cref{sec:example_fss} is adopted to illustrate the principle of \gls{SJDE} procedures and to underpin the differences to \gls{JDE} methods that use a fixed sample size.
As this model does not fulfill the requirements for the \gls{ao} procedure, it is not included.
A detailed presentation of the \gls{ao} procedure and its comparison to the optimal one is given in \cite{reinhard2024}.
The aim is to design a procedure that used on average as few samples as possible, while the Type-I and Type-II error probabilities should not exceed $0.05$ and the \gls{MSE} for the parameter under the alternative should not exceed $0.2$.
Moreover, the maximum number of samples is restricted to $\Nmax=100$.
For the used model, the sample mean $\bar{x}_n$ serves as sufficient statistic.

\begin{figure}
     \centering
     \includegraphics{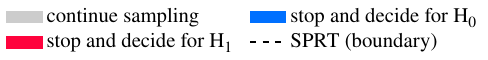}

     \includegraphics{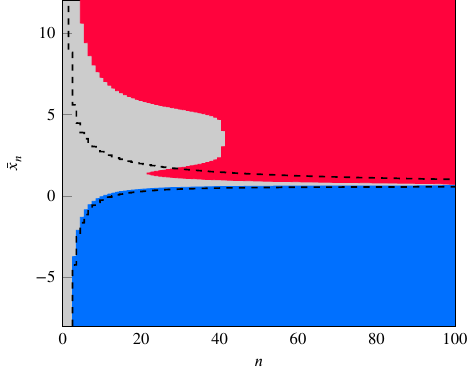}\caption{Policy of the optimal sequential procedure: Region in which the procedure stops sampling (\textcolor{contColor}{\rule{0.6cm}{0.2cm}}), region in which the procedure stops and accepts the null hypothesis(\textcolor{decColorH1}{\rule{0.6cm}{0.2cm}}), and regions in which the procedure stops and accepts the alternative(\textcolor{decColorH2}{\rule{0.6cm}{0.2cm}}). The boundary of the \gls{sprt} policy  is indicated by a dashed line.}
     \label{fig:opt_policy}
\end{figure}
The optimal policy and the boundary of the \gls{sprt} policy are illustrated in \cref{fig:opt_policy}.
First, the optimal policy is inspected.
As can be seen, sampling is stopped directly if the sample mean is a large negative number as this leads to a high confidence that the null hypothesis is true.
On the other hand, if the sample mean is a large positive number, the confidence that the alternative is true is high along with a small uncertainty about the true parameter.
Hence, sampling is stopped early in this case.
Though a sample mean close to zero leads to a small uncertainty about the true parameter, a sample mean close to zero does not indicate a preference for either of the two hypotheses.
Therefore, a sample mean close to zero results in taking an additional sample.
If the sample mean lies in $1\leq\bar{x}_n<5$, more samples are required before the procedure stops sampling, which is due to high uncertainty about the true parameter.
This is caused by the fact that the cost of accepting a particular hypothesis, and, hence, also the cost for stopping, is a linear combination of a penalty term for wrong decisions and a penalty term for inaccurate estimates.
The weights of the individual terms, also referred to as cost coefficients, are then chosen such that the constraints are fulfilled by solving \cref{eq:seq_opt_coeff}.
In summary, one can see that the fact whether to stop or to continue sampling depends on the information about the underlying hypothesis and the unknown parameter contained in the observations.

Contrary to this, the policy of the \gls{sprt} shows a different behavior.
For a negative sample mean, which corresponds to a decision in favor of the null hypothesis, the policy of the \gls{sprt} is similar to the optimal policy for \gls{SJDE}, which is due to the fact that there is no estimation if a decision in favor of the null hypothesis is made.
For a positive sample mean, i.e., the region that corresponds to a decision in favor of the alternative, the \gls{sprt} policy differs significantly from the optimal one.
Especially, the region $1\leq\bar{x}_n<5$ in this the optimal procedure continues sampling due to the uncertainty about the true parameter does not exist in case of the \gls{sprt}.
Moreover, it can be seen that the corridor around $\bar{x}n=0$ in which the procedures continue sampling is much wider for the \gls{sprt} than for the optimal procedure for \gls{SJDE}.

\begin{table}
     \centering
     \caption{Simulation results. Comparison of Type-I and Type-II error probabilities ($\detErr[0]$ and $\detErr[1]$), the MSE ($\estErr[1]$) and the average sample number of an optimal procedure (opt) in the sequential and \gls{fss} scenario as well as a sequential two-step procedure (TS). For the \gls{fss} optimal procedure, the number of samples has been chosen such that all constraints are fulfilled.}
     \label{tbl:sim_res_sequential}
\begin{tabular}{lcccc}
     \toprule
     & & \multicolumn{2}{c}{Sequential} & FSS \\
     \cmidrule(lr){3-4}
     Metric & Target & Opt. & TS & Opt. \\
     \midrule
     $\detErr[0]$ & 0.050 & 0.049 & 0.028 & 0.050 \\
     $\detErr[1]$ & 0.050 & 0.050 & 0.039 & 0.050 \\
     $\estErr[1]$ & 0.200 & 0.199 & 0.930 & 0.197 \\
\midrule
     $\E[\tau]$             & -- & 22.10 & 12.15 & 36.00 \\
     $\E[\tau\given\Hyp_0]$ & -- & 11.40 & 14.41 & 36.00 \\
     $\E[\tau\given\Hyp_1]$ & -- & 32.78 &  \phantom{0}9.88 & 36.00 \\
     \bottomrule
     \end{tabular}     
\end{table}

Next, we compare the performance of the optimal sequential procedure with the targeted error levels, the performance of a sequential two-step procedure and the performance of an optimal \gls{fss} procedure.
The sequential two-step procedure comprises an \gls{sprt} with threshold chosen by Wald's approximation followed by an \gls{mmse} estimator.
For the optimal \gls{fss} procedure, the sample size is chosen such that it fulfills the constraints on the error levels.
The Monte Carlo results are summarized in \cref{tbl:sim_res_sequential}.
First, it can be seen that the optimal sequential procedure hits the targeted error levels exactly.
Contrary to this, the sequential two-step procedure has much smaller error probabilities than the targeted ones, while the constraint on the \gls{MSE} is severely violated.
Additionally, the sequential two-step procedure uses on average much fewer samples than the optimal sequential ones.
Both are caused by the fact that stopping rule of the sequential two-step procedure is entirely determined by the sequential detector, i.e., the confidence about the true hypothesis, and uncertainty about the true parameter is not considered at all.
As the sequential two-step procedure ignores the uncertainty about the true parameter, it uses, given that the alternative is true, on average much fewer samples.
From the last column in \cref{tbl:sim_res_sequential}, one can see that the \gls{fss} procedure needs significantly more samples to achieve a similar performance as the optimal sequential procedure for joint detection and estimation in terms of error probabilities and \gls{MSE}.

In summary, the major advantages of optimal sequential procedures for joint detection and estimation are: (i) due to the additional degree of freedom introduced by the variable sample number, all performance measures, i.e., the error probabilities and the \gls{MSE}, can be controlled individually  and (ii) it uses, on average, significantly fewer samples than its \gls{fss} counterpart with similar error levels

 \section{Open Problems and Future Directions}\label{sec:open_problems}
This section covers open problems and future research directions in (sequential) joint detection and estimation.
First, we present open problems that stem from existing research and briefly discuss possible solution strategies.
Next, we discuss the application of \gls{SJDE} in the context of changepoint detection and distributed sensor networks.

\subsection{Performance Bounds for (Sequential) Joint Detection and Estimation}
Deriving performance bounds is an essential topic in statistical signal processing.
A widely used performance bound in parameter estimation is the \gls{CRB} that provides a lower bound on the variance for an unbiased estimator.
Besides the original formulation, there exist also Bayesian versions \cite{trees2007bayesian}, versions in the case of model mismatch \cite{fortunati2016misspecified,fortunati2017performance} and sequential versions \cite{wolfowitz1947efficiency,simons1980sequential,Ghosh01011987}.
Moreover, a modification on the \gls{CRB} exist for the scenario in which a detection step is present prior to estimation, see, e.g., \cite{chaumette2005influence}.
In that work, the expectation when calculating the Fisher information matrix is conditioned on the event that a target is detection and the result is referred to as conditional Fisher information matrix.
Although this could be used to derive a bound for the \gls{JDE}, it does highly depend on how the estimation errors are quantified, see the discussion in \cref{sec:jde}.
Moreover, it only considers the \gls{fss} scenario.
Especially in the sequential scenario, strictly optimal procedures are often hard to design.
As one is often interested in designing lightweight procedures that are close to optimal, a lower bound on the \gls{MSE} in the case of \gls{SJDE} would allow quantifying the performance gap to strictly optimal procedures. 
Depending on the problem at hand, the latter might even not yet be implementable.

\subsection{Robust (Sequential) Joint Detection and Estimation}
In statistical robustness one aims to develop inference methods that can tolerate a certain amount of deviation from the assumed model without breaking down.
Especially the assumption of normality, which is typically imposed on the noise processes, is often violated in practice \cite{zoubir2012robust}.
The field of robust statistics dates back to the groundbreaking work of Huber \cite{huber1964robust_estimation,huber1965robust_detection} in the 1960s.
Since then, robust detectors and robust estimators for a fixed number of samples have been developed.
See, for example, \cite{fauss2021minimax} and \cite{zoubir2012robust,zoubir2018robust} for an overview on robust detection and estimation, respectively.
Moreover, robust methods for the problem of sequential detection and robust sequential change detection have been investigated \cite{vos2001minimax,kharin2002robustifying,fauss2020minimax,xie2022minimax,yang2024sequential}.
The amount of existing works on robust detection and robust estimation underpins the necessity of statistical robustness.
However, the work on robust (sequential) joint detection and estimation is rather scarce.
The problem of \gls{JDE} under prior uncertainty is addressed by a restricted Bayes approach in \cite{bayram2016joint,dulek2018restricted}.
Recently, a first attempt to the analysis and design of minimax optimal procedures for joint detection and estimation under uncertainty of the likelihood has been made in \cite{reinhard2026minimax}.
In \cite{reinhard2016approach}, an approach to robustify a procedure for sequential joint detection and estimation has been proposed and sequential joint detection and estimation in distributed sensor networks under non-Gaussian noise has been addressed in \cite{reinhard2021distributed}.

In \gls{JDE}, one usually has a rather complex statistical model that leaves plenty of room for model uncertainty.
Theoretically, one could model the uncertainty of the joint distribution of the data, the parameters and the hypothesis, which will be most probably not feasible in practice.
In the sequential scenario, the time dependency of the distributions makes it even more complicated.
Hence, realistic and tractable uncertainty models for the problem of (sequential) \gls{JDE} have to be investigated as a first step.
Deepening the theory of robust \gls{JDE} and, subsequently, transferring these result to the sequential case should be a next step.
However, as the design of strictly minimax procedures is often challenging, cf. \cite{fauss2021minimax,reinhard2026minimax}, the derivation of an almost minimax solution is important from a practical perspective.
Although these methods lose robustness over strictly minimax ones, they are often much easier to implement.
Therefore, they provide a good compromise between robustness and practicability.

\subsection{Sequential Joint Detection and Estimation for Complex Data Structures}
The design of optimal sequential procedures as outlined in \cref{sec:opt_seq_jde} builds on a self-contained framework.
However, as the optimal procedure is characterized by the recursively defined cost function $\rho_n(\stat)$, the implementation suffers from the curse of dimensionality.
To overcome this, several different directions exist.
The first one is the derivation of procedures that become optimal in some asymptotic regime\footnote{Here, we avoid the term \emph{asymptotically optimal} as existing literature distinguishes between \emph{asymptotically optimal} and \emph{asymptotically pointwise optimal} procedures. See, for example, for both definitions \cite{bickel1968asymptotically}.}, i.e., procedures that become optimal when the targeted error levels tend to zero and perform close to optimal for moderate targeted error levels.
This is a widely used method in sequential analysis, see, e.g., \cite{kiefer1963asymptotically,bickel1967asymptotically,bickel1968asymptotically,draglia1999multihypothesis,cohen2015asymptotically,he2021asymptotically}, and has already been applied to sequential joint detection and estimation \cite{reinhard2024,reinhard2021}.
However, \cite{reinhard2024} has the major drawback that a parameter to be estimated is required under both hypotheses and, therefore, does not support the widely used model in which the nominal state, i.e., the model under the null hypothesis, is completely known.
Here, the focus of future research should be to derive \gls{ao} procedures for the case when there is no estimation under a particular hypothesis or in which the estimation errors are not quantified by the \gls{MSE}.
Besides \gls{ao} procedures, it is possible to exploit machine learning techniques to approximate the optimal solution.
Since the cost function $\rho_n(\stat)$ is a recursively defined Bellman equation, one could use (deep) reinforcement methods \cite{sutton2020,Arulkumaran2017,Li2023} to learn the optimal policy.
The last direction to overcome the curse of dimensionality is called deep optimal stopping.
Here, deep neural networks approximate the optimal stopping rule that is usually characterized by the non-linear Bellman equation.
Such methods have been successfully applied to high-dimensional optimal stopping problems in finance.
For more details, see, for example, \cite{
becker2019deep,becker2021solving,damera2023deep} and references therein. 
All three research direction to tackle the optimal stopping problem are equally important to apply sequential joint detection and estimation to more complex data structures.

Even when the optimal stopping rule can be derived or well approximated, the design, i.e., the search for the optimal cost coefficients, can be quite challenging.
For small state spaces of $\stat$, the optimal coefficients can be efficiently found by means of linear programming \cite{fauss2015linear, reinhard2018,reinhard2022}.
However, for the \gls{ao} procedure or if the optimal stopping rule are learned, the design reduces to a gradient-based search, such as a projected gradient ascent \cite{reinhard2019a} or a \gls{bfgs} approach \cite{reinhard2022}, that often relies on Monte Carlo simulations.
Evaluating the objective and the gradients are both very costly as both rely on Monte Carlo simulation, and possibly a learning step to learn the new stopping rule, if learning techniques are used.
Due to this high computational load, classical step-size control such as the conditions by Wolfe’s or Armijo’s seem to be too conservative and would result in too many iterations.
To mitigate this effect, a first custom step-size control has been published in \cite{reinhard2022}.
Additionally, the number of iterations highly depends on the initial values of the cost coefficients.
By using results from the \gls{fss} scenario and by asymptotic theory, a set of initial value could be approximated.
Therefore, future research should also focus on the development of fast and powerful optimization algorithms tailored to the problem of sequential joint detection and estimation.

\subsection{Sequential Joint Changepoint Detection and Parameter Estimation}
The online detection of changes in the statistical model of streaming data is an important problem in statistics and signal processing with a wide range of applications including industrial process monitoring \cite{oakland2007statistical}, power system monitoring \cite{Chen2016Quickest}, communications \cite{Lai2008} or sensor networks \cite{halme2026multi}.
Here, the aim is to detect a change as quickly as possible while controlling the false alarm rate.
More details on (sequential) change detection can be found, e.g., in \cite{poor2009quickest,tartakovsky2014sequential,xie2021}.
Though the statistical model in the pre-change regime is assumed to be known, the statistical model in the post-change regime often depends on unknown parameters.
To solve this problem, existing methods include an estimation step, although its outcome is not of primary interest, but it can rather be seen as an auxiliary step towards detection, see, e.g.,  \cite{Lai1998,xie2023window,Halme2025}.
However, we believe that one is often also interested in the estimate of the post-change parameters and, hence, controlling the quality of the estimate is an important aspect.
Therefore, the question of deriving an (asymptotically) optimal stopping rule that controls the false alarm rate and the estimation quality is worth investigating.
The results on \gls{fss} as well as on sequential \gls{JDE} obtained so far should give a good starting point for addressing sequential joint changepoint detection and parameter estimation.

\subsection{Sequential Joint Detection and Estimation in Sensor Networks}
In many modern applications, a variety of spatially dispersed sensors that form some kind of sensor network are used.
To end up with scalable and fault-tolerant inference algorithms, the sensors, also referred to as nodes or agents, should exchange their data with their neighboring sensors and process the data locally rather than transmitting everything to a central processing unit.
When the data is assumed to be streaming data, i.e., it is observed sequentially, \gls{ci} \cite{Kar2013} like algorithms are preferable to solve a task in a collaborative manner.
The main idea of these algorithms can be described as follows.
At every time instant, the state of a node is first averaged with the states of neighboring nodes (\emph{consensus}) and then updated by the average of state updates, that are based on the current samples, of the neighboring nodes (\emph{innovations}).
Such algorithms have been successfully applied to the problem of sequential detection \cite{Sahu2016,Leonard2018,Li2018} and sequential joint detection and estimation \cite{reinhard2020b,reinhard2021distributed}.
However, for the latter, it has been assumed that the data is (almost) Gaussian distributed.
To extend previous work to a broader class of distributions, such as the ones from the exponential family, the following coupled questions arise. (i) how can the state be defined, (ii) how can the state be aggregated, (iii) how can the aggregated states be characterized statistically.
Distributed networks also come with new challenges in the context of statistical robustness, such as the existence of Byzantine nodes \cite{Li2021}. \section{Conclusions}
In this work, we have provided a comprehensive overview of joint detection and estimation, covering both suboptimal and optimal procedures. 
In addition to the conceptual introduction, we have discussed how advanced problem formulations affect the presented theory.
Additionally, after a short introduction on sequential analysis, we have presented suboptimal and optimal sequential procedures for joint detection and estimation.
Using a numerical example, we have shown the advantages and power of sequential methods over counterparts that use a fixed number of samples.
Finally, we have addressed open problems and discussed future research directions.

\bibliographystyle{elsarticle-num}
\bibliography{references}

\vskip3pc         

\bio{author_dr}
Dominik~Reinhard received his B.Sc., M.Sc. and Dr.-Ing. from Technische Universit\"at Darmstadt, Germany, all in electrical engineering and information technology in 2012, 2016 and 2021, respectively.
In May 2016, he joined the Signal Processing Group at Technische Universit\"at Darmstadt, where he successfully defended his Ph.D. thesis in August 2021.
Subsequently, he worked as a Postdoctoral Research Fellow at the Signal Processing Group at Technische Universit\"at Darmstadt until March 2026.
 
 His current research interests are in statistical signal processing with special emphasis on detection and estimation theory, sequential analysis and statistical robustness. 
\endbio

\vskip3pc         

\bio{author_az}
{Abdelhak~M.~Zoubir} is an IEEE Life Fellow, a Fellow of EURASIP, and an IEEE Distinguished Lecturer (Class 2010-2011). He holds since 2003 the position of Professor at Darmstadt University of Technology, Germany, and was Chair of Signal Processing and Head of the Signal Processing Group until March 2026. His research interest lies in statistical methods for signal processing with emphasis on bootstrap techniques, robust detection and estimation and array processing applied to telecommunications, radar, sonar, car engine monitoring and biomedicine. He published over 500 journal and conference papers and two books on these areas. Professor Zoubir was General Chair, Technical Program Chair and Member of Organization Committees of numerous international conferences and workshops, most notably he was Technical Program Co-Chair of the largest and flagship conference on signal processing, The International Conference on Acoustics, Speech, and Signal Processing (ICASSP), held in Florence in 2014. He served on numerous editorial boards, most notably, he was the Editor-In-Chief of the IEEE Signal Processing Magazine (2012-2014), the flagship publication of the IEEE Signal Processing Society (SPS). Dr Zoubir was elected as Chair (2010-2011) of the IEEE SPS Technical Committee Signal Processing Theory and Methods (SPTM. He served on the Board of Governors of the IEEE SPS as elected Member-at-Large (2015-2017), and elected Member of the Board of Directors (BoD) of the European Association for Signal Processing (EURASIP) from 2009-2016 and was its President from 2017 until 2018. He has been inducted to the German National Academy of Science and Engineering (acatech) in January 2024.
\endbio

\end{document}